\documentclass[aps,amsmath,amssymb,superscriptaddress,onecolumn]{revtex4} 
\usepackage{graphicx}
\usepackage{psfrag}
\DeclareMathOperator{\Tr}{Tr}
\DeclareMathOperator{\MyRe}{Re}

\DeclareMathOperator{\Prob}{Prob}
\numberwithin{equation}{section}
\renewcommand{\theequation}{\arabic{section}.\arabic{equation}}
\begin{document}
 
\title{Quantum Noise Analysis of Spin Systems Realized with Cold Atoms}
\author{Robert W. Cherng and Eugene Demler}
\affiliation{Department of Physics, Harvard University, 
Cambridge, Massachusetts 02138, USA}
\date{\today}

\begin{abstract}
We consider the use of quantum noise to characterize many-body states
of spin systems realized with ultracold atomic systems.  These systems
offer a wealth of experimental techniques for realizing strongly interacting
many-body states in a regime with a large but not macroscropic number of atoms where
fluctuations of an observable such as the magnetization 
are discernable compared to the mean value.  The full distribution 
function is experimentally relevant and encodes high order correlation functions 
that may distinguish various many-body states.  We apply quantum noise analysis 
to the Ising model in a transverse field and find a distinctive
even versus odd splitting in the distribution function for the transverse magnetization
that distinguishes between the ordered, critical, and disordered phases.
We also discuss experimental issues relevant for applying quantum noise analysis
for general spin systems and the specific results obtained for the Ising model.
\end{abstract}
 
\maketitle

\section{\label{sec:intro}Introduction }

\subsection{Strongly correlated systems of cold atoms}

A natural regime for ultracold akali gases is when interactions are weak.
For typical experimental densities of $n \approx 10^{14}$ atoms per
cubic centimeter and scattering lengths of a few nanometers,
interparticle distances are two orders of magnitude larger than the
scattering length. This regime was the focus of most experimental
work done in the first decade since the observation of Bose-Einstein
condensation \cite{boulder-95,mit-95,rice-95}. 
Excitement in this field was fueled by the possibility of
turning {\it gedankenexperiments} into reality.
This includes interference of
independent condensates \cite{mit-97a}, 
creation of atom lasers \cite{mit-97a,mit-97b},
and condensation of multicomponent systems
\cite{mit-98a,mit-98b,jila-98,hamburg-04,gatech-04,tokyo-04,berkeley-05}.  
The Gross-Pitaevski equation captures all essential
features of systems in this regime \cite{smith-book,stringari-book}. 
This is a mean-field 
approximation that
only requires the knowledge of a single wavefunction describing a
macroscopically occupied state describing the condensate. 
Conceptually, the many-body aspect is important only in turning 
quantum coherence into a macroscopic phenomenon.
However, solutions to the non-linear equations may be non-trivial
with the vortex lattice \cite{ho-01,fetter-03,baym-04a} being
a prime example.

The current focus for the field of ultracold atoms is 
shifting towards creation and study of systems with
strong interactions and correlations. The driving force of this
evolution is remarkable experimental progress which provide powerful
tools for realizing strongly interacting systems of cold
atoms. Magnetically tuned Feshbach resonances
(see \cite{stoof-04,koehler-06} and references therein)
can be used to make
scattering lengths comparable to interparticle distances. This allowed
creation of molecules and observation of pairing in fermionic
systems. Confining particles in optical lattices is another
powerful tool for amplifying the role of interactions.  Atoms in a
periodic potential created by a standing wave laser beam can exhibit
both enhanced interactions and suppressed kinetic energy. The relative
strength of the two energies can be tuned by changing the intensity of
the laser beam. This technique was used to observe the superfluid to
Mott transition in which the behavior of atoms changed from wavelike
in the superfluid phase to particle like in the Mott phase \cite{greiner-02}. 
Reduced dimensionality is another useful tool for 
making strongly correlated systems.
The strength of an effective one-dimensional interaction can be changed
by tuning either the particle density, transverse confinement, or 
scattering length.
This approach was used in several experiments that
explored the crossover from weakly interacting quasicondensate to the
Tonks gas of hard core bosons \cite{weiss-04,paredes-04}. There are many
other promising techinques for obtaining strongly interacting systems
of quantum degenerate particles including rotating condensates \cite{cornell-99},
ultracold polar molecules (see \cite{doyle-04} and references therein),
and Rydberg atoms \cite{rydberg-99,rydberg-00}.

One of the primary motivations for studying strongly correlated
systems of cold atoms is the expectation that they will realize
important and not yet fully understood quantum many body states.  The
target list of modern experiments includes the 
Fulde-Ferrel-Larkin-Ovchinnikov phase of
unbalanced fermion mixtures with attraction \cite{mit-06,rice-06}, supersolid
phase \cite{lewenstein-02,buchler-03,sarma-03}, 
quantum magnetic phases of atoms in optical lattices
(ordered or spin liquids) \cite{demler-03,cirac-04,zoller-06,lewenstein-06a}, fermions with d-wave pairing in
lattice systems with repulsive interactions \cite{hofstetter-02}, and quantum Hall
phases \cite{gunn-99,cooper-99,regnault-04}. An important open question is detection and
characterization of these states.  

Analysis of correlation functions is a common way of characterizing
many-body states in condensed matter physics.  Neutron and light
scattering experiments can be used to measure spin and charge response
functions.  Spin and charge density wave orders give rise to
additional $\delta$-function peaks in the static correlation
functions, hence they immediately show up as new Bragg peaks in
elastic neutron scattering experiments. Conductivity measurements at
finite frequency (microwave, infrared, optical, etc.) provide access
to current-current correlation functions \cite{mahan-book}. They are a
useful tool for detecting states with a gap in the quasiparticle
spectrum, such as superconducting and various density wave
states. Photoemission experiments measure electron spectral functions
and can be used to study excitation spectrum or observe electron
fractionalization in one dimensional systems
\cite{kivelson-01,kivelson-02,annarbor-03}.  So what we want
for strongly correlated systems of cold atoms is to have a tool box
for measuring correlation functions. 

Certain experimental techniques used in cold atoms provide natural
analogues of condensed matter experiments. For example, Bragg
spectroscopy is similar to light and neutron scattering in solid state
physics and can be used to analyze the dynamic response functions
\cite{mit-01}. RF spectroscopy in cold atom systems is analogous to
finite frequency conductivity measurements \cite{rf-03a,rf-03b,rf-04a,rf-04b}. 
There are also several
proposals for designing experiments that mimic solid state
methods \cite{levin-06,baym-04b}. Additional techniques such as 
time of flight and interference experiments \cite{boulder-95,mit-95}
or the possibility of single atom detection
\cite{esslinger-05,aspect-05}
are also available which are not feasible with traditional condensed matter systems.
The richness of physical
phenomena that we expect from strongly correlated systems of cold
atoms provide strong motivation to think about new methods and
approaches that would provide ``smoking gun'' signatures of various
many-body phases.

\subsection{From noise analysis to correlation 
functions in interacting cold atoms systems}

The usual philosophy of performing measurements in physics is to 
minimize the noise by averaging over many experiments.
However, there are  notable exceptions. In mesoscopic solid
state physics and quantum optics, one is often interested in the noise
rather than the average signal. For example, shot noise was used to
measure fractional charge and statistics of excitations in
fractional quantum Hall systems \cite{fqhe-97a,fqhe-97b}. Quantum noise in photon
counting experiments was used to
identify non-classical states of light \cite{qo-book}.  

The philosophy of noise analysis was implemented in several recent types of experiments
with cold atoms.  These experiments analyzed quantum noise in time of flight images
of atoms released from an optical lattice \cite{hadzibabic-04,foelling-05,aspect-05,greiner-05,porto-06}
(see Ref. \cite{altman-04} for theoretical analysis). 
Free expansion during the time of flight maps
momentum distribution of atoms inside the interacting system into the
density distribution after the expansion. 
Correlation functions $\langle \rho(r) \rho(r') \rangle$ of the images in 
real space contains information about the $ \langle n_k n_{k'}
\rangle$ correlation function of the interacting system in momentum space. 
Note that $n_k = c_k^\dagger c_k$ is the occupation number of momentum states
and not the density operator at wavevector $k$. Hence $\langle n_k
n_{k'} \rangle$ represents a non-local correlation function for the
original interacting system. 

What can one learn from $ \langle n_k n_{k'} \rangle$ correlation functions?
The simple answer is the periodiciy of the original system.
Quasimomentum (i.e. lattice momentum) is a good quantum number in a
periodic system. Quasimomentum differs from the physical momentum by Bragg
reflections by reciprocal lattice vectors. This leads bunching for bosons
and peaks in $ \langle n_k n_{k'} \rangle$ when the two momenta differ
by a reciprocal lattice vector $k-k'=G$. Analogously, fermions exhibit 
antibunching.  Such experiments have been done for spinless
bosons \cite{hadzibabic-04,foelling-05,porto-06} as well as fermions
\cite{greiner-05} and provided 
one of the most elegant realizations of Hanbury-Brown-Twiss experiments with neutral
atoms. When such experiments are performed with mixtures, we can
expect even more exciting results. They should provide an ideal probe
for antiferromagnetism or static density wave, since both phenomena
increase the size of the unit cell and give rise to additional
reciprocal lattice vectors \cite{altman-04}. It is worth pointing out that
fluctuating orders, like in Luttinger liquids, can also be
observed using this method \cite{mathey-05}.
Another beautiful application of the noise analysis in time of flight
experiments was the measurement of fermion pairing on the BEC side of
the Feshbach resonance \cite{greiner-05}. These experiments demonstrated correlations in
the occupation numbers for fermions with momenta $k$ and $-k$, which
demonstrated the existence of a Cooper pair condensate at zero momentum.

A different method for analyzing correlation functions in interacting
many-body systems is based on interference experiments between two
independent condensates.  It may not be immediately obvious why this
measurement corresponds to the noise analysis. In every shot we should
find an intereference pattern. However the position of interference
fringes is unpredictable from shot to shot. If we average over many
experiments we will find that the interference pattern washes out
completely. We can put this in a more formal mathematical language:
Let $A$ be the quantum mechanical operator that describes the
amplitude of interference fringes \footnote{The spatial profile of the
density measured by the absorption image is $\rho(x) = 2A_0+A e^{iQx}
+ A^* e^{-iQx}$. Here $A$ is the amplitude of interference
fringes. Contrast is defined as $A/A_0$}, we have $\langle A \rangle
=0$ but finite $\langle |A|^2 \rangle $.  What is interesting for our
purposes is the value of $\langle |A|^2 \rangle $ itself, or the
related quantity of the fringe contrast. When individual condensates
are not fluctuating, we expect fringes with perfect contrast. When
there are strong thermal or quantum fluctuations we expect suppression
of the fringe amplitude. Theoretical analysis performed in
Ref. \cite{polkovnikov-06} showed that the amplitude of interference
fringes can be related to the instantaneous two point correlation
function: $\langle |A|^2 \rangle = \Omega \int_\Omega G^2(r)$, where
$G(r)=\langle c(r) c^\dagger(0) \rangle$ and $\Omega$ is the imaging
area.  By analyzing scaling of $\langle |A|^2 \rangle $ with the size
of the images region, Hadzibabic et al. recently demonstrated the 
Berezinskii-Kosterlitz-Thouless
transition in two dimensional condensates \cite{hadzibabic-06}.

The common philosophy of HBT experiments with atoms 
\cite{hadzibabic-04,foelling-05,aspect-05,
greiner-05,porto-06,altman-04,mathey-05,polkovnikov-06,hadzibabic-06} and
analysis of the average contrast in interference of independent
condensates (IIC) experiments is the focus on second order coherence.
This means that in the analysis of absorption images, we are not
looking for peaks in $\langle \rho(r) \rangle$ but in $\langle \rho(r)
\rho(r') \rangle$ \footnote{We remind the readers that quantum
mechanical averaging implies averaging over many measurements}.  An
interesting question to ask is whether one can use quantum noise
measurements to analyze higher order coherence. In the case of IIC
experiments this problem was considered in Refs.
\cite{polkovnikov-06,gritsev-06}. The basic idea is that we do not only
measure the average value of the amplitude of interference fringes but
also their fluctuations. And it is
straightforward to show that that higher moments of $|A|^2$ correspond
to higher order correlation functions. In the case of one dimensional
Bose systems with interactions it is possible to find the explicit
form of the distribution function of $|A|^2$.  For weakly interacting
bosons phase fluctuations within condensates are small and one finds
the Gumbel distribution of $|A|^2$ with a small width. When
interactions between atoms become so strong that the system approaches
the limit of hard core bosons, the distribution function becomes
Poissonian.

\subsection{Quantum magnetism with cold atoms in optical lattices}

The main subject of this paper is to extend the approach of quantum
noise analysis to the study of higher order correlations in quantum
magnetic systems realized using cold atoms. Several approaches have
been discussed recently for creating magnetic systems with cold atoms
(see Ref. \cite{lewenstein-06a} for a review). Our proposal is
sufficiently general so we expect that it can be applicable to all of
them. The main idea can be understood in the following
example. Consider a measurement of the $\alpha$ component of spin
magnetization in a one dimensional spin systems
\begin{eqnarray}
\label{eq:magnetization}
M_\alpha = \sum_i S_\alpha(i).
\end{eqnarray}
The average magnetization can be obtained by averaging over many
experiments
\begin{eqnarray}
\langle M_\alpha \rangle = \sum_i \, \langle S_\alpha(i) \rangle
\end{eqnarray}
but individual measurements will exhibit quantum noise. Each
measurement will give a value of $M_\alpha$ that is different from the
average $\langle M_\alpha \rangle$. It is easy to see that the strength
of quantum fluctuations is controlled by correlation functions of
the spin operators. For example
\begin{eqnarray}
\langle M^2_\alpha \rangle = \sum_{i_1,i_2} \, \langle S_\alpha(i_1) S_\alpha(i_2)
\rangle
\end{eqnarray}
So the width of the distribution  is determined by two point
correlation functions. By the same argument higher moments provide a
measure of higher order correlation functions
\begin{eqnarray}
\langle M^{n}_\alpha \rangle = \sum_{i_1 \dots i_n} \, \langle S_\alpha(i_1)
\dots S_\alpha(i_n) \rangle
\end{eqnarray}
The knowledge of high moments is contained in the distribution
function of $M_\alpha$. So if we know the distribution function of $P(M_\alpha)$
we obtain information about correlation functions of arbitrarily
high order. In real experiments, of course, technical noise and a
finite number of experiments will limit the number of moments that
can be  reliably extracted.

Generalization of this idea to other measurements is straightforward:
one should measures not just the average value of some operator but
its entire distribution function. The distribution function contains
information about high moments which can be
related to high order correlation functions. This provides valuable
characterization of a many-body state.  It is important that
measurements are done on a mesoscopic system, hence fluctuations will
not be neglibly small compared to the mean value. For macroscopic
solid state samples this is not easy to achieve, but for cold atoms
systems with tens or hundreds of thousands of atoms this is an
experimentally relevant regime. Another advantage of the cold atoms
systems is their nearly complete isolation from the environment. We
point out that Ref. \cite{bruder-05} considered a related idea of
measuring the number of particles in a mesoscopic part of a fermionic
system with pairing. They showed that fluctuation in the total number
of particles will exhibit interesting evolution as one tunes the
system from the BCS to the BEC regime.

In the main part of this paper contained in Sec. \ref{sec:distribution}
we consider a specific example of the
magnetization distribution for the Ising model in a transverse field. We
are not the first ones to address this question.
Ref. \cite{eisler-03} study the distribution functions for both
the longitudinal and transverse magnetizations of this model with
analytical and numerical techniques.  We focus primarily on the transverse
magnetization but provide a more detailed analysis of its global structure.
In particular, we study how higher order moments encode the different correlations 
in the ordered, critical, and disordered phases and give a simple physical picture
in terms of confinement versus proliferation of domain walls.
Sec. \ref{sec:exp}
discusses experimental issues involved in applying quantum noise analysis
to ultracold atoms including those specific to the results we obtained
for the transverse Ising model.  We summarize our discussion in the
conclusion of Sec. \ref{sec:conclusion}.  The appendices contain
mathematical details of the discussion in Sec. \ref{sec:distribution}.

\section{\label{sec:distribution}Magnetization distribution in the Transverse Ising Model}

The one-dimensional Ising model in a transverse field
provides a textbook example of a symmetry-breaking quantum phase transition \cite{sachdev-99}.
It is described by the Hamiltonian
\begin{equation}
\label{eq:spin_hamiltonian}
H=-J\sum_{i=1}^N\left[2S_x(i)S_x(i+1)+gS_z(i)\right]
\end{equation}
where $S_\alpha(i)$ are spin-$1/2$ operators at site $i$. 
Here $g$ is a dimensionless coupling which sets the strength of the transverse field and $J$
sets the overall energy scale.
We take periodic boundary conditions and $N$ is the number of sites.
The competition between the transverse field and the Ising term of the Hamiltonian 
drives a quantum phase transition at $g_c=1$ from an 
ordered phase $g<g_c$ with spontaneous magnetization along the $x$ direction to a disordered phase $g>g_c$ with 
zero magnetization along $x$.
We are interested in the distribution function for the magnetization $M_\alpha$ given by
\begin{equation}
P(M_\alpha)=\langle \delta(M_\alpha) \rangle
\end{equation}
where $\delta(x)$ is the delta function and brackets denote ground state expectation values.

The global structure of $P(M_x)$ describing the 
order parameter follows from low-order moments of the longitudinal magnetization
$M_x$ and symmetry breaking arguments.
Away from the critical point $g\ne g_c$, exact results on $\langle S_x(i)S_x(j)\rangle$ 
\cite{pfeuty-70,barouch-71a} show the spin correlation length and the second moment
$\langle M_x^2\rangle$ are finite.
Physically, this means spins fluctuate independently on length scales
larger than the correlation length.
Thus, for a sum of a large number of spins, the corresponding distribution should
be approximately gaussian.
In the broken symmetry phase $g<g_c$, the distribution should have two peaks 
describing the formation of spontaneous magnetization for $M_x$.
On the other hand, the unbroken symmetry phase $g>g_c$ should only have one peak
centered at zero.
At the critical point $g=g_c$, the power-law decay of 
$\langle S_x(i)S_x(j)\rangle\sim\sum |i-j|^{-1/4}$ gives a divergent contribution to 
$\langle M_x^2\rangle$ and 
a non-trivial distribution function.

More analysis is needed for the transverse magnetization $M_z$ where symmetry breaking
arguments do not apply.
Exact results \cite{pfeuty-70,barouch-70} for $\langle S_z(i)S_z(j)\rangle$
show the second moment $\langle M_z^2\rangle$ is finite even at the critical point. 
Analogously, $P(M_z)$ is also gaussian.  However, additional aspects of the global structure
such as the number of peaks and other features are encoded in higher order
moments.

Instead of analyzing individual moments, we study the entire distribution
$P(M_z)$ directly.
We find its global structure reflects the amount of 
\textit{disorder} in the system.
This is in contrast to $P(M_x)$ which reflects the amount of \textit{order}.
We show $P(M_z)$ is asymptotically gaussian for a large number of spins
with a characteristic splitting between even and odd magnetization 
that differs between the phases.
A duality mapping shows this behavior has a physical interpretation in terms of 
proliferation, criticality, and confinement of dual domain walls when 
$g<g_c$, $g=g_c$, and  $g>g_c$, respectively.  The domain wall physics
is particularly clear in the small and large $g$ regimes where $P(M_z)$ describes
the qualitatively different counting statistcs of pairs and single domain walls, respectively.
In section \ref{sec:calculation} we calculate $P(M_z)$ via fermionization.  
We obtain and discuss its global structure in section \ref{sec:global},
focusing on the even versus odd splitting and its interpretation using a duality mapping.
Appendix \ref{app:toeplitz} discusses the use of Toeplitz determinants 
employed in the calcuation.  In addition, appendix \ref{app:g} derives
the small and large $g$ behavior in more detail.  Finally,
Appendix \ref{app:saddlepoint} describes the saddle-point analysis
used to obtain the asymptotic gaussian behavior.

\subsection{\label{sec:calculation}$P(M_z)$ Distribution Function}
Jordan-Wigner fermionization (see Ref. \cite{lsm-61}) maps
spins in one dimension onto spinless fermions
\begin{eqnarray}
\label{eq:jw}
c_i&=&\prod_{j<i}[-\sigma_z(j)]\sigma_-(i)\\
c^\dagger_i&=&\prod_{j<i}[-\sigma_z(i)]\sigma_+(i)
\end{eqnarray}
where $\sigma_\alpha(i)=2S_\alpha(i)$ are the Pauli matrices
and $\sigma_\pm=\sigma_x\pm i\sigma_y$ are the corresponding raising and lowering operators.
The spin Hamiltonian of Eq. \ref{eq:spin_hamiltonian} maps to free fermions
\begin{equation}
\label{eq:fermion_hamiltonian}
H=-J\sum_{k>0}\left[(g+\cos k)(c^\dagger_k c_k+c^\dagger_{-k} c_{-k})
-i\sin k(c_{-k}c_k-c^\dagger_k c^\dagger_{-k})\right]
\end{equation}
where $c_k$ are in momentum space.

Instead of studying $M_z$ directly, the analysis is simpler for the quantity
\begin{equation}
\label{eq:rhon}
\rho_n=\sum_{i=1}^{n}\left(\frac{1}{2}-S_z(i)\right)=\sum_{i=1}^{n}\left(1-c^\dagger_i c_i\right)
\end{equation}
which is simply related to both $M_z$ and the fermion number.
Recall $N$ is the number of spins in the entire chain and compare this
to $n$ which is the number of spins in a block embedded in this system.
We will take $N\rightarrow\infty$ first to describe the thermodynamic limit
and then $n\rightarrow\infty$.  Physically, we are studying the local magnetization
$M_z$ of a large block of spins embedded in a much larger system.  This has important
consequences in the interpretation of the results and experimental detection.

We will calculate the generating function for $\rho_n$ defined as
\begin{equation}
\label{eq:chin}
\chi_n(\lambda)=\langle e^{i\lambda \rho_n} \rangle
\end{equation}
where the brackets denote ground state expectation values in
the thermodynamic limit $N\rightarrow\infty$.
The corresponding distribution function is given by the fourier transform of the
generating function
\begin{equation}
\label{eq:pn}
P_n(m)=\int_{-\pi}^{\pi}\frac{d\lambda}{2\pi}e^{-i\lambda m}\chi_n(\lambda)
\end{equation}
which is related to $P(M_z)$ by the following
\begin{equation}
P(M_z)=P_n(n/2-M_z).
\end{equation}

We first define the following linear combinations of fermionic operators
\begin{eqnarray}
\label{eq:fermion_combination}
A_i&=&c_i+c_i^\dagger\\
B_i(\lambda)&=&c_i+e^{i\lambda}c_i^\dagger
\end{eqnarray}
and obtain for the generating function
\begin{equation}
\label{eq:chin_fermion}
\chi_n(\lambda)=\langle A_1 B_1(\lambda)\ldots A_n B_n(\lambda)\rangle
\end{equation}
where we have used the operator identity 
$e^{i\lambda (1-c_i^\dagger c_i)}=A_i B_i(\lambda)$.
Using Wick's theorem, the expectation value in Eq. \ref{eq:chin_fermion}
can be reduced to products of the following two-point functions 
\begin{eqnarray}
\langle A_i A_j \rangle &=& \int_{-\pi}^{\pi}\frac{dk}{2\pi}e^{-ik(i-j)}\\
\langle B_i(\lambda) B_j(\lambda)\rangle &=&
\int_{-\pi}^{\pi}\frac{dk}{2\pi}e^{-ik(i-j)}
e^{-2i\lambda}\left[1+\sin(\lambda)\sin(\theta_k)\right]\\
\label{eq:ab}
\langle A_i B_j(\lambda)\rangle &=&
\int_{-\pi}^{\pi}\frac{dk}{2\pi}e^{-ik(i-j)}
\frac{1}{2}\left[(1+e^{i\lambda})+
(1-e^{i\lambda})e^{i\theta_k}\right]
\end{eqnarray}
where the Bogoliubov angle 
\begin{equation}
\label{eq:bogoliubov}
e^{i\theta_k}=\frac{g+e^{ik}}{\sqrt{(g+e^{ik})(g+e^{-ik})}}
\end{equation}
comes from diagonalizing the Hamiltonian in Eq. \ref{eq:fermion_hamiltonian}
using a Bogoliubov transformation.

Notice $\langle A_i A_j \rangle=\delta_{ij}$.  However, the contractions of this
form entering $\chi_n(\lambda)$ vanish because they 
always occur with $i\ne j$.  Moreover, any full contraction containing
$\langle B_i B_j \rangle$ also vanishes since it must contain some
$\langle A_i A_j\rangle$ contraction.  This implies the only 
full contractions that contribute are those that only pair $A_i$ with $B_j$.
Taking into account the fermionic signs,
these full contractions organize into a determinant
of the form
\begin{equation}
\label{eq:chin_f}
\chi_n(\lambda)=\det_{i,j=1\ldots n}\left[f_{i-j}(e^{i\lambda})\right]
\end{equation}
where $f_n(z)$ are the Fourier coefficients of the function
\begin{equation}
\label{eq:f}
f(z,x)=\sum_{n=-\infty}^{\infty}f_n(z)x^n=
\frac{1}{2}\left[(1+z)+(1-z)\frac{g+x}{\sqrt{(g+x)(g+x^{-1})}}\right].
\end{equation}
Notice that the coefficients are given by
$f_n(e^{i\lambda})=\langle A_i B_j(\lambda)\rangle$
and the function $f(z,x)$ is the integrand in Eq. \ref{eq:ab} with
$z=e^{i\lambda}$ and $x=e^{ik}$.

The above result gives an explicit expression for the 
generating function.  Here we can
make some general comments on the structure of $\chi_n(\lambda)$.
By diagonalizing the matrix inside the determinant, we see that
the generating function is of the form
\begin{equation}
\chi_n(\lambda)=\prod_{i=1}^n
\frac{1}{2}\left[(1+e^{i\lambda})+(1-e^{i\lambda})\mu_i \right].
\end{equation}
Here $\mu_i$ are the eigenvalues of the $n\times n$ matrix $M_{ij}$ formed from
the Fourier coefficients of the Bogoliubov angle
\begin{equation}
M_{ij}=\int_{-\pi}^{\pi}\frac{dk}{2\pi}e^{-ik(i-j)}e^{i\theta_k}.
\end{equation}
This can be rewritten in the more suggestive form
\begin{eqnarray}
\chi_n(\lambda)&=&\prod_{i=1}^n\chi_B\left(\lambda,p_i\right)\\
\chi_B(\lambda,p)&=&(1-p)+pe^{i\lambda}\\
p_i&=&\frac{1}{2}(1-\mu_i)
\end{eqnarray}
as a product of generating functions $\chi_B(\lambda,p)$.
We recognize these generating functions as that of a Bernoulli random variable 
with probability $p$ of obtaining one and $1-p$ of obtaining zero.
This implies $\rho_n$ is the sum of independent Bernoulli variables.
This is not surprising since $\rho_n$ is directly related to the fermion number and 
each fermionic level can be either occupied or unoccupied.

However, even though the matrix $M_{ij}$ is real, the eigenvalues $\mu_i$ are only required to
occur in complex conjugate pairs.  
In fact, we find the $\mu_i$ are generically complex and satisfy 
$|\mu_n|\le 1$.  This means that $\chi_B(\lambda,p_i)$ cannot be interpreted as a 
classical Bernoulli distribution.
However, a pair of conjugate eigenvalues gives
\begin{equation}
\label{eq:chin_bernoulli_pair}
\chi_B(\lambda,p_i) \chi_B(\lambda,p_i^*)=
\frac{1}{4}\left[|1-\mu_i|^2+2(1-|\mu_i|^2)e^{i\lambda}+|1+\mu_i|^2e^{2i\lambda}\right]
\end{equation}
allowing an  interpretation as a classical distribution with the coefficient
of $e^{im\lambda}$ giving the probability of obtaining $m$.
Notice that the interference of the complex eigenvalues $\mu_i$ can lead
to suppression of odd versus even values.
In the fermionic description, this is a result of BCS-type pairing in the Hamiltonian
of Eq. \ref{eq:fermion_hamiltonian}.  The anomalous term $a_k a_{-k}$ 
and its hermitian conjugate describe pair creation and annhilation processes
which might favor even occupation.
We note that the appearance
of independent Bernoulli variables for the fermion number distribution
has been studied in the mathematical literature (see Ref. \cite{hough-06})
as determinantal point processes.  However, it appears the
cases that have been studied correspond to states with no BCS-type pairing and 
purely real eigenvalues $\mu_i$.  
These heuristic considerations take into account
only one or two fermionic levels.  Determining the global structure of $P_n(m)$
is a many-body problem involving a large number of fermionic levels which we address
in the next section.

\subsection{\label{sec:global}Global structure}

Before discussing the global structure of $P(M_z)$,
we first argue that it gives a measure
of the amount of \textit{disorder} in the system.  
Physically this is not surprising since 
$S_z(i)$ tends to disorder spins along the $x$ direction by inducing tunneling
between the two polarizations.
This competes with the Ising term $S_x(i)S_x(i+1)$ which favors 
ordering of spins along the $x$ direction.
A duality transformation \cite{savit-80} makes this connection more concrete.  
This transformation maps the spin Hamiltonain of Eq. \ref{eq:spin_hamiltonian} to 
a spin Hamiltonian of the same form but with the coupling constant exchanged
\begin{equation}
\label{eq:dual_hamiltonian}
H=-J\sum_{i=1}^N\left[2gT_x(i)T_x(i+1)+T_z(i)\right].
\end{equation}

The $T_\alpha(i)$ operators describe dual spins that obey the same commutation 
relations as $S_\alpha(i)$ and are given by 
\begin{eqnarray}
\label{eq:duality}
T_z(i)&=&2S_x(i-1)S_x(i)\\
T_x(i)&=&\frac{1}{2}\prod_{j\le i}[2S_z(i)].
\end{eqnarray}
in terms of the $S_\alpha(i)$ operators.
Under this transformation the operator $\rho_n$ maps to 
\begin{equation}
\label{eq:rhon_dual}
\rho_n=\sum_{i=1}^{n}\left(\frac{1}{2}-2T_x(i-1)T_x(i)\right)
\end{equation}
where the term under the summation has eigenvalue 0 for two dual spins aligned
on the $x$ direction and eigenvalue 0 for anti-aligned.
This implies that $\rho_n$ is the number operator for dual domain walls
and $P_n(m)$ is its probability distribution.

In particular, the fermionic operators $c_i$ ($c^\dagger_i$) create (annhilate) dual domain walls.
For the BCS-type Hamiltonian of Eq. \ref{eq:fermion_hamiltonian}, 
$Jg$ acts as the chemical potential for free fermions with dispersion $-J\cos(k)$ with
the anomalous terms describing pair creation and annhilation.
When $g$ is small, the chemical potential crosses the band and pair fluctuations
cost little energy leading to dual domain wall proliferation.
When $g$ is large, the chemical potential is below the bottom of the band and 
dual domain walls are confined due to energy cost.  
Note this is the exact opposite of the physical domain walls for the $S_x(i)$
spins but the underlying physics of confinement versus proliferation \cite{fradkin-78} is the same.

Analysis of $P_n(m)$ shows how the distribution function reflects the above physics
Numerically, Eq. \ref{eq:chin_f} allows for efficient calculation of 
the distribution function $p_n(m)$.  The determinant
representation gives $\chi_n(\lambda)$ as a polynomial of degree $n$ in 
$e^{i\lambda}$ and the coefficient of $e^{im\lambda}$ can be extracted
via a computer algebra system to give $P_n(m)$.
Analytical insight is also possible in the large-$n$ limit through the use
of Toeplitz determinant asymptotics outlined in Appendix \ref{app:toeplitz}.  
We find that
the generating function has the following asymptotic behavior
\begin{equation}
\label{eq:chin_asymptotic}
\chi_n(\lambda)\sim\exp\left[c_0(e^{i\lambda}) n + 
c_1(e^{i\lambda}) \log n +c_2(e^{i\lambda})\right]
\end{equation}
where the coefficients $c_i(z)$ are explicitly known functions.

\begin{figure}
\begin{center}
\psfrag{x}[c][c]{$m$}
\psfrag{y}[c][c]{$P_n(m)$}
\psfrag{g=0.5}[c][c]{$g=0.5$}
\psfrag{g=2.0}[c][c]{$g=2.0$}
\includegraphics[width=3in]{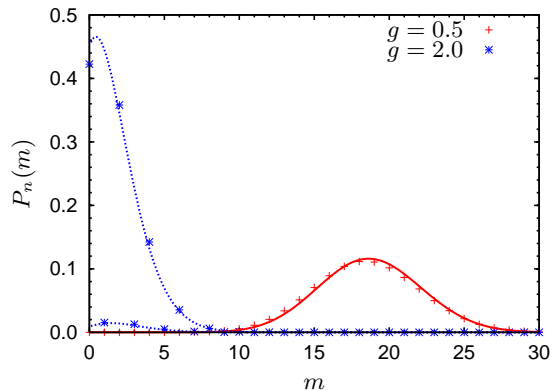}
\caption{Distribution function $P_n(m)$ for $g=2$, $g=0.5$ and $n=50$.  Solid lines denote
the small $g$ binomial distribution and dashed lines denote the large $g$ poissonian distribution.
Notice the separate distributions for even and odd $m$ for $g>g_c$.  The
center and width of the peak tend to zero for $g\rightarrow \infty$ but
approach a constant for $g\rightarrow 0$.
}
\label{fig:compare}
\end{center}
\end{figure}

We first consider the small and large $g$ regimes.
We carry out the expansions in Appendix \ref{app:g} and find for the generating 
function
\begin{equation}
\label{eq:chin_g}
\chi_n(\lambda)\sim
\begin{cases}
\left(\frac{2+g}{4}+\frac{2-g}{4}e^{i\lambda}\right)^n & g\ll g_c\\
\exp\left[\frac{n-1}{16 g^2}(e^{2i\lambda}-1)+\frac{1}{8 g^2}(e^{i\lambda}-1)\right]& g\gg g_c
\end{cases}
\end{equation}
where we recognize the form of the generating functions as that of a binomial random variable
for $g\ll g_c$ and a sum of two Poissonian random variables for $g\gg g_c$.
However, notice in the $g\gg g_c$ regime that one poissonian variable depends on $e^{2i\lambda}$ 
whereas the other depends on $e^{i\lambda}$.  
This implies the operator $\rho_n$ behaves as 
\begin{equation}
\label{eq:rhon_g}
\rho_n \sim \begin{cases}
B\left(n,\frac{2-g}{4}\right)&g\ll g_c\\
2 P\left(\frac{n-1}{16g^2}\right)+P\left(\frac{1}{8g^2}\right)&g\gg g_c
\end{cases}
\end{equation}
where $B(n,p)$ is a binomial random variable with $n$ trials and probability
of success $p$ and $P(\lambda)$ is a Poissonian random variable with parameter $\lambda$.
Explicit expressions for $P_n(m)$ for these two regimes are given in
Appendix \ref{app:g}.
We plot the distribution $P_n(m)$ for $g=0.5$ and $g=2$ in Fig. \ref{fig:compare}.
Notice the distinct splitting between even and odd values of $m$ for $g<g_c$ 
that is absent for $g>g_c$.

\begin{figure}
\begin{center}
\begin{tabular}[b]{cc}
\includegraphics[width=3in]{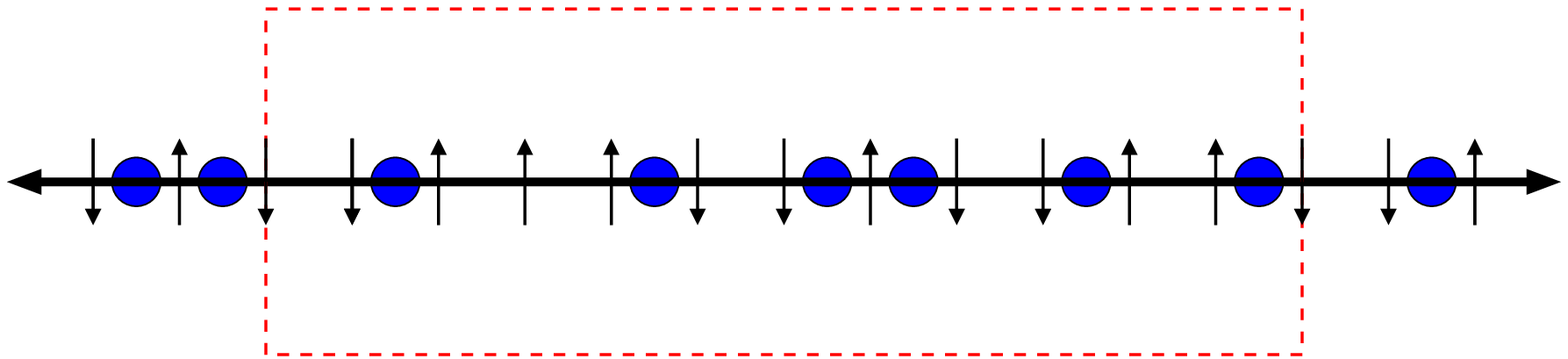}&\textbf{(a)}\\
\includegraphics[width=3in]{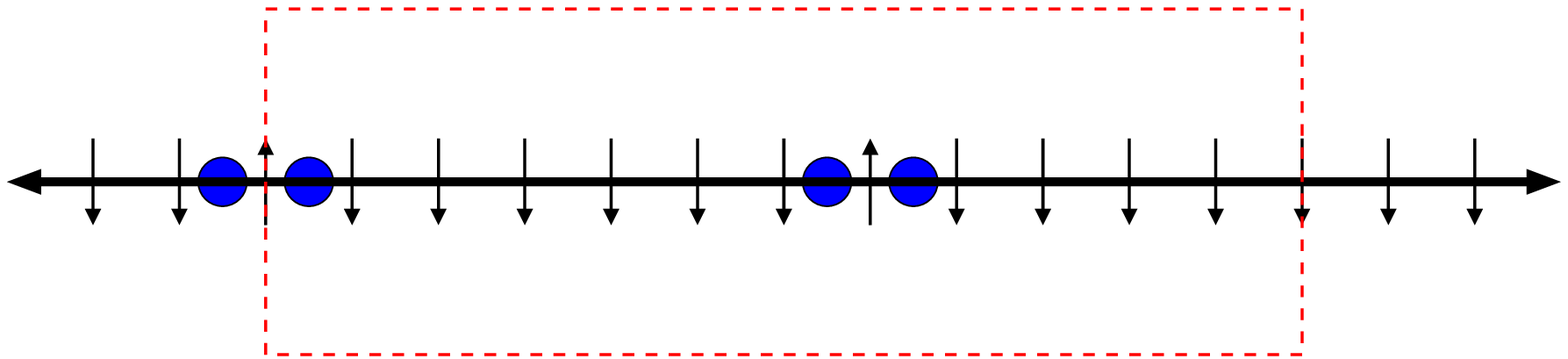}&\textbf{(b)}
\end{tabular}
\caption{
Calculating $P(M_z)$ for a block of spins (red rectangle) maps to counting 
dual domain walls (blue circles) within.
In the proliferated phase for $g<g_c$ \textbf{(a)}, \textit{individual} domain walls
are free whereas in the confined phase \textbf{(b)}, \textit{pairs} of domain walls are 
free giving qualitatively different statistics.  Notice pairs in the bulk
and at the boundary contribute differently in the confined phase.
The small arrows indicate the corresponding dual spin configuration.
}
\label{fig:dw}
\end{center}
\end{figure}

The qualitatively different behavior has a simple physical interpretation in terms
of proliferation versus confinement of dual domain walls.
We have argued $M_z$ simply counts the number of dual domain walls.  When they proliferate 
for small $g$, individual domain walls are free.
Each site contributes independently with probability $P_{occ}=1/2-g/4$ 
of being occupied and $P_{empty}=1-P_{occ}$ of being empty.
This leads to the binomial counting statistics that does not distinguish between
even and odd total occupation.
In contrast, the confined regime for $g>g_c$ has dual domain walls tightly bound into pairs that occur
with low probability $P_{pair}=1/16g^2$.
In the dilute limit, the counting statistics of pairs is Poissonian.  However, pairs in the bulk
contribute a factor of two to the number of dual domain walls with a probability $(n-1)P_{pair}$ that scales
with $n$ whereas pairs at the boundary contribute a factor of one to the total number with a probability
$2P_{pair}$ since there are two boundaries.  The differing contriubtions of pairs in the bulk and boundary
leads to the splitting between even and odd total occupation.  A schematic illustration of counting
dual domain walls in these two regimes is illustrated in Fig. \ref{fig:dw}.

The above analysis relies on an expansion in the small parameter $g$ or $g^{-1}$
and identifying the resulting form of $\chi_n(\lambda)$.
In order to study the evolution from the small to large $g$ regimes, we use a large-$n$
saddle-point integration for $P_n(m)$ which works for all $g$.
This comes at the expense of obtaining only the asymptotic gaussian behavior
and is carried out in Appendix \ref{app:saddlepoint}.
The gaussian nature of the resulting distribution can be established two ways.  
We have already argued that $\rho_n$ is
the sum of $n$ independent Bernoulli-like random variables with finite variance
which means $P_n(m)$ is guassian by the central limit theorem.  In addition,
we can examine the Taylor series
\begin{equation}
\log \chi_n(\lambda)=\sum_{p=0}^{\infty} \frac{\lambda^p}{p!}\langle \rho_n^p\rangle_c.
\end{equation}
whose coefficients $\langle \rho_n^p \rangle_c$ give the cumulants or connected moments.
We find all cumulants are finite and proportional to $n$ at leading order.  
By considering the ratio
\begin{equation}
\label{eq:gaussian}
\frac{\langle \rho_n^p \rangle_c}{\langle \rho_n^2 \rangle_c^{p/2}}\sim n^{1-p/2}
\end{equation}
we conclude that the first two cumulants completely characterize the distribution
for large-$n$ which implies that $P_n(m)$ is gaussian.  This is not
in contradiction to the small and large $g$ regimes since the resulting distributions 
are also gaussian for sufficiently large $n$.

\begin{figure}
\begin{center}
\psfrag{g}[c][c]{$g/g_c$}
\psfrag{m}[c][c]{$\mu$}
\psfrag{s}[c][c]{$\sigma$}
\includegraphics[width=3in]{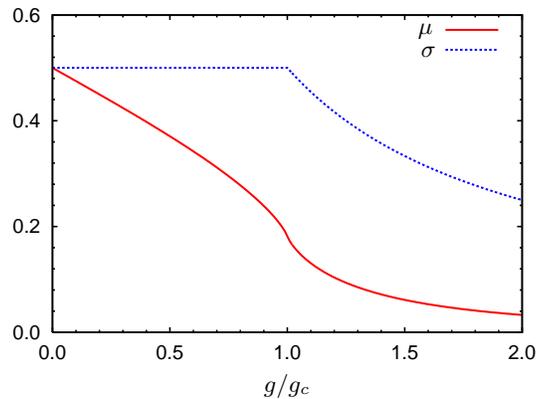}
\caption{Mean $\mu$ and standard deviation $\sigma$ as a function of $g/g_c$ for 
the asymptotic gaussian form of $P_n(m)$.  Notice $\mu,\sigma\rightarrow 1/2$ for
$g\rightarrow 0$ and $\mu,\sigma\rightarrow 0$ for $g\rightarrow \infty$
}
\label{fig:param}
\end{center}
\end{figure}

The absence or presence of an even versus odd splitting
characterizes the small and large $g$ regimes, respectively.
This aspect of the global structure of $P_n(m)$ reflects higher order
moments which encode the underlying correlations of the model and 
appear in the large-$n$ gaussian distributions.
In particular, we find 
\begin{equation}
\label{eq:pn_gaussian}
P_n(m)\sim \frac{e^{-(m-n\mu)^2/2n\sigma^2}}
{\sqrt{2\pi n \sigma^2}}
\begin{cases}
1+(-1)^m(1-g^{-2})^{1/4}&g>g_c\\
1+(-1)^m C n^{-1/4}& g=g_c\\
1&g<g_c
\end{cases}
\end{equation}
which is of the form of either a single gaussian or the sum of two gaussians.
The $(-1)^m$ term gives the even versus odd splitting which has a sharp transition
and critical behavior at $g=g_c$.  This illustrates clearly
that the quantum phase transition manifests itself in higher order correlations
of $M_z$ and is accessible via the distribution function $P(M_z)$.

Having disucssed the physical origin of the this splitting,
we now show how it also reflects correlations of the order parameter or
longitudinal spins $S_x(i)$.
At first this is suprising since $P_n(m)$ describes correlations of the
transverse spins $S_z(i)$.  However, recall the $S_z(i)$ correspond to domain walls
for $T_x(i)$  which are related to $S_x(i)$ via duality.
Explicitly, we obtain From Eq. \ref{eq:rhon_dual} 
\begin{equation}
\exp(i\pi \rho_n)=4T_x(0)T_x(n).
\end{equation}
This gives the two-point function of the dual spins $T_x(0)T_x(n)$ as a 
function of $\rho_n$ implying that $P_n(m)$ can be used to compute its average
in the ground state
\begin{equation}
\label{eq:tautau_identity}
\langle T_x(0)T_x(n) \rangle = \frac{1}{4}\sum_{m} P_n(m) (-1)^m.
\end{equation}
The interpreation of this equation is also illustrated in Fig. \ref{fig:dw}.
An even (odd) number of dual domain walls between two sites gives a spin configuration
with spins aligned (anti-aligned) along the $x$ direction.
By using the gaussian form of $P_n(m)$ in Eq. \ref{eq:pn_gaussian}
and passing from summation to integration we find
\begin{equation}
\label{eq:tautau_result}
\langle T_x(0)T_x(n) \rangle\sim
\frac{1}{4}\begin{cases}
(1-g^{-2})^{1/4}&g>g_c\\
C n^{-1/4}& g=g_c\\
0
\end{cases}
\end{equation}
where we have dropped the contribution from terms entering with $(-1)^m$ in
the summation.  Here $C=2^{1/12}e^{1/4}A^{-13}$ with $A$ being Glaisher's constant.
This agrees with known results on the spin correlation function $S_x(0)S_x(n)$
\cite{pfeuty-70,barouch-71a} after applying
a duality transformation $g\rightarrow g^{-1}$.  
Why the agreement is exact is discussed in Appendix \ref{app:saddlepoint}.

\begin{figure}
\begin{tabular}[b]{cc}
\psfrag{x}[c][c]{$x$}
\psfrag{y}[c][c]{$P(x)dx$}
\includegraphics[width=3in]{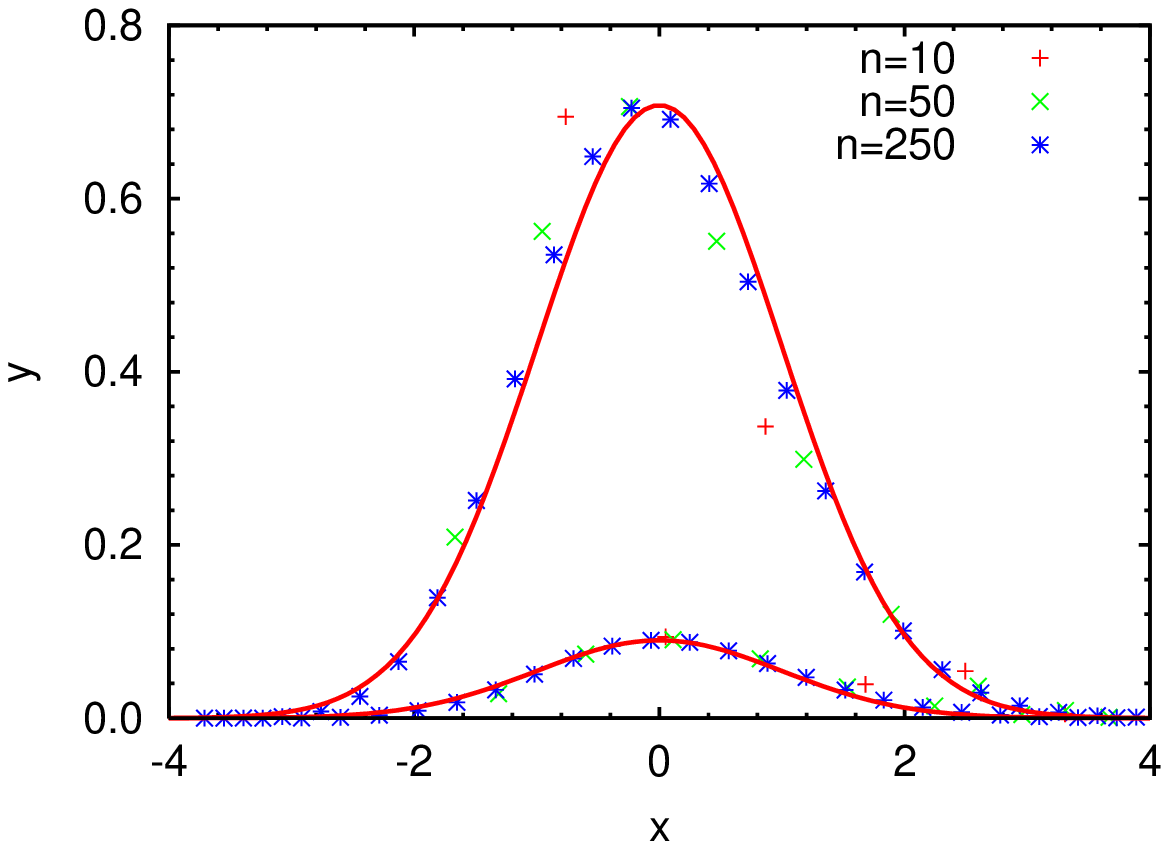}&\textbf{(a)}\\
\psfrag{x}[c][c]{$x$}
\psfrag{y}[c][c]{$P(x)dx$}
\includegraphics[width=3in]{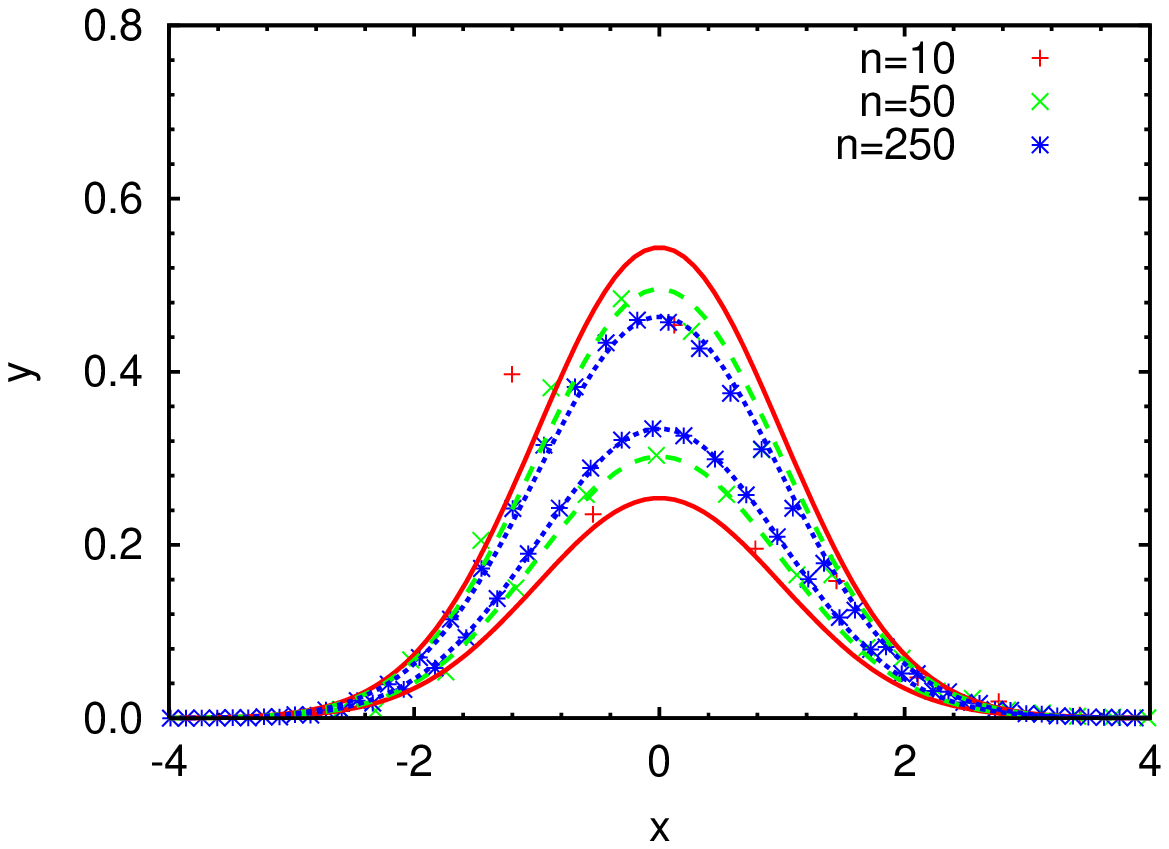}&\textbf{(b)}\\
\psfrag{x}[c][c]{$x$}
\psfrag{y}[c][c]{$P(x)dx$}
\includegraphics[width=3in]{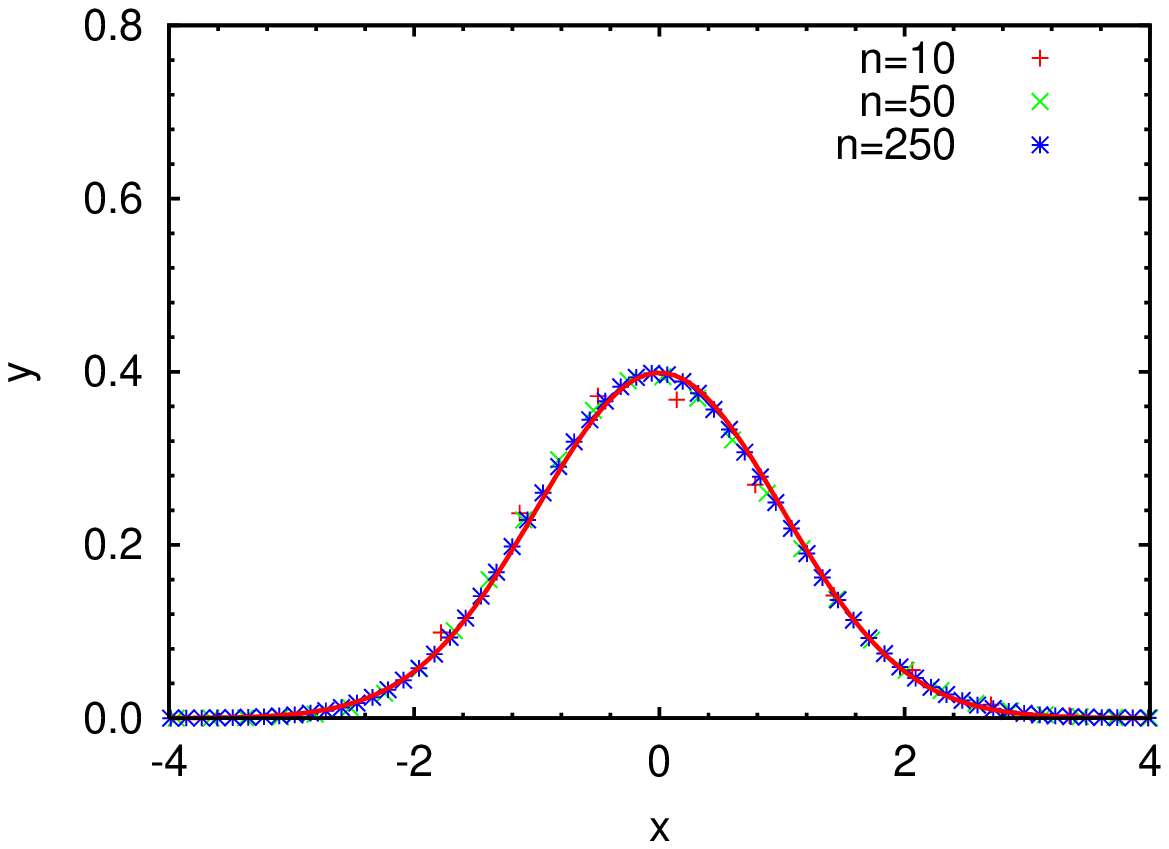}&\textbf{(c)}
\end{tabular}
\caption{Distribution function $P_n(m)$ for $g=1.25$ \textbf{(a)}, $g=g_c=1$ \textbf{(b)},
and $g=0.8$ \textbf{(c)} plotted as a function of the rescaled variables 
$x=(m-\langle \rho_n\rangle_c)/\langle\rho_n^2\rangle_c^{1/2}$ and 
$P(x)dx=P_n(m)\langle\rho_n^2\rangle_c^{1/2}$ where $\langle\ldots\rangle_c$ denotes the
cumulant.  The solid lines indicate the asymptotic gaussian form of $p_n(m)$.
Notice the convergence to guassian behavior even at $g=g_c$ and the splitting between
even and odd $m$ for $g\ge g_c$.
}
\label{fig:distribution}
\end{figure}

We now discuss the mean $\mu$ and variance $\sigma^2$ of the guassian distributions
in Eq. \ref{eq:pn_gaussian}.
They are expressed in terms of the Bogoliubov angle by
\begin{eqnarray}
\label{eq:pn_parameters}
\mu&=&\frac{1}{2}\int_{-\pi}^{\pi}\frac{dk}{2\pi}\left(1-e^{i\theta_k}\right)\\
\sigma^2&=&\frac{1}{4}\int_{-\pi}^{\pi}\frac{dk}{2\pi}\left(1-e^{2i\theta_k}\right).
\end{eqnarray}
We plot the mean $\mu$ and standard deviation $\sigma$ in Fig. \ref{fig:param}.
As $g\rightarrow 0$, the distribution has a finite width $\sigma$ and a finite mean $\mu$.
In contrast, as $g\rightarrow\infty$, both $\sigma$ and $\mu$ go to zero.  This 
behavior can also be seen in Fig. \ref{fig:compare}.  We plot $P_n(m)$
in rescaled variables in Fig. \ref{fig:distribution}.
The asymptotic convergence to gaussian distributions is quite
good even at the critical point $g=g_c$ for sufficiently large $n$.
In addition, the distinctive even versus odd splitting is absent for $g<g_c$,
present for $g>g_c$, and decays algebraically for $g=g_c$.

Here we compare our results with that of Ref. \cite{eisler-03}.
Our new contribution is the demonstration and analysis of the even versus odd splitting
in $P_n(m)$.  Fig. \ref{fig:dw} shows that this is fundamentally a boundary 
effect of a local block of spins.  Thus it is necessary
to separate the system size $N$ from the block size $n$ to see this effect.
In contrast, Ref. \cite{eisler-03} implicitly set $N=n$ in the beginning of their
analysis.  At a more technical level, we find an approach based on mapping the
problem to Toeplitz determinants allows for separation of the length scales $N$ and $n$.
This allows us to obtain the first finite size correction to the leading
gaussian behavior to $P_n(m)$ as discussed in Appendix \ref{app:toeplitz}
which turns out to be crucial in obtaining the splitting.
In addition, this approach also facilitates study of the small and large $g$
regimes where the underlying physics is most evident.
As another technical note, Ref. \cite{eisler-03} also obtains the 
asymptotic gaussian behavior via a saddle-point analysis, but we show
that there is an additional saddle-point
which gives rise to the splitting and is discussed in Appendix \ref{app:saddlepoint}.
Having discussed one application of quantum noise analysis in a specific spin system,
we now turn our attention to a discussion of general issues in experimental realizations.

\section{\label{sec:exp}Experimental Issues}
Although the use of ultracold atoms in experimental studies of quantum many-body systems 
has seen great progress, techniques based on quantum noise analysis are just
becoming within reach.  The fundamental difficulties arise from the need
to collect accurate statistics on noise fluctuations of an observable,
ideally to high orders.  Here we discuss some proposals for experiments
and the relevant issues for quantum noise analysis.

Optical lattices offer one of the most promising avenues for engineering
many-body systems in ultracold atoms including spin systems 
\cite{lewenstein-06a}
but they may pose problems 
for noise analysis.  This is especially true for low-dimensional systems.    
They have already been realized in optical lattices
\cite{paredes-04,weiss-04,hadzibabic-04} but occur in a large array of
up to thousands of one- or two-dimensional systems within each sample.  
A single measurement such as an absorption image over the entire sample 
corresponds essentially to averaging over the large array which
supresses fluctuations.
Low-order statistics such as second-order
coherence \cite{hadzibabic-04,foelling-05,porto-06} are still accessible.
However
higher order noise fluctuations may be difficult to detect as a result
of the intrinsic averaging that occurs within each sample.

Accessing higher order noise may be possible through
realizations of single or small numbers of low-dimensional systems
where fluctuations are not negligble compared to the average.
Such an approach has already been used to study the 
Berezinskii-Kosterlitz-Thouless transition \cite{hadzibabic-06}.
In this experiment, the quantum noise in the interference amplitude of
just a pair of two-dimensional quasi-condensates was used to detect signatures
of the transition. 
For one-dimensional systems, surface microtraps 
allow for tight confinement in a waveguide geometry 
\cite{thywissen-00,grangier-01,zimmermann-01}
and optical box traps have recently demonstrated single 
one-dimensional condensates \cite{singlesite-05}.

In addition to preparation, experiments will also have to
advance in measurement techniques.  The ideal measurement of
quantum noise for magnetization would be state-selective
\textit{in situ} imaging with 
single atom precision to obtain accurate high order statistics
and high resolution to resolve spatial correlations.
This has not yet been achieved with
optical absorption imaging but may be possible through
proposed scanning tunneling microscropy of ultracold
atoms \cite{gericke-06}.  However, single atom counting has 
been demonstrated in an atom laser setup \cite{esslinger-05}.
However, this is more suited to the study of temporal correlations
rather than the spatial correlations describing noise in
magnetization.
Detection of single atoms  has also been used in Hanbury-Brown-Twiss
experiments \cite{aspect-05} but with imaging after expansion
which resolves momentum correlations and not \textit{in situ} 
which probes real space correlations.
An alternative measurement involves imprinting the quantum
noise for magnetization onto a weakly coupled laser probe
\cite{lewenstein-06b}.  In this case, the quantum noise of
the laser itself should be less than that of the magnetization
and may make features at the single atom level difficult to
detect.

The above discussion of general issues for quantum noise experiments
also applies to the transverse Ising model we considered.  However,
resolving the even versus odd splitting in the distribution function 
requires additional considerations.
As shown in Fig. \ref{fig:dw}, distinguishing between even and odd 
magnetization requires a sharp boundary between the block
of spins undergoing measurement and the rest of the spin chain.
This requirement does not pose difficulties for measurement via
high resolution \textit{in situ} imaging.  However,
for techniques that couple to the entire sample,
RF cutting (see Ref.\cite{rfcut-96}) may be needed to isolate the 
region of interest beforehand. 

Sensitivity at the single atom level 
will also be crucial 
for detecting the even versus odd splitting.
We consider the feasibility of such a measurement in the
absence of \textit{in situ} imaging by analyzing 
the proposal of
Ref. \cite{lewenstein-06b} for coupling the magnetization
to off-resonant coherent laser light.
For our purposes, the main result relates the output quadrature
of light $X_{out}$ to the input quadrature $X_{in}$
and the magnetization $M_z$ as
\begin{equation}
\label{eq:quadrature}
X_{out}=X_{in}-\frac{\kappa}{\sqrt{n_a}}M_z
\end{equation}
where $\kappa$ is the atom-photon interaction and 
$n_a$ is the number of atoms.
Here the input quadrature component $X_in$ 
has been rescaled by the
number of photons $n_p$ and describes coherent light 
with noise $\Delta X_{in}=1/\sqrt{2}$

The quantum noise of the magnetization is thus imprinted
on top of the inherent photon shot noise of light.
Discerning gross features of the magnetization distribution
only requires fluctuations in $X_{out}$
due to quantum noise in the \textit{collective} magnetization 
be greater than fluctuations due to photon shot noise.  The width
of the magnetization distribution gives an estimate
$\Delta M_z \sim \sqrt{n_a}\sigma$ where $\sigma$ is shown
in Fig. \ref{fig:param} and yields the condition
\begin{equation}
\label{eq:condition1}
\kappa>\frac{1}{\sigma\sqrt{2}}.
\end{equation}
This enables resolving the smooth distribution shown in Fig. 
\ref{fig:compare}  in the regime with no splitting whereas it 
only gives the average of the two separate distributions of
the same figure in the regime with splitting.
Detecting this splitting requires $X_{out}$ fluctuations
due to \textit{single} spins rise above photon shot noise
and yields a much more stringent condition
\begin{equation}
\label{eq:condition2}
\kappa>\sqrt{\frac{n_a}{2}}.
\end{equation}

Eq. \ref{eq:quadrature} describing the coupling of magnetization to 
light is valid when decoherence due to spontaneous emission is small
and for low photon absorption.  This implies the atomic depumping
$\eta=n_p\sigma_{r}\Gamma^2/A\Delta^2$ and photon absorption 
$\epsilon=n_a\sigma_{r}\Gamma^2/A\Delta^2$ are small.
Here $n_p$ is the number of photons, $\sigma_{r}$ is the resonant
scattering cross section which is distint from $\sigma$ the
width of the magnetization distribution, $\Gamma$ is the spontaneous decay rate,
$A$ is the illuminated cross sectional area, and $\Delta$ is
the detuning.
The atom-photon interaction is given by 
$\kappa=\sqrt{n_a n_p}\sigma_{r}\Gamma/A\Delta$.
For given $\eta,\epsilon\ll 1$, the condition for resolving
collective magnetization noise in Eq. \ref{eq:condition1} becomes
\begin{equation}
\frac{\Gamma}{\Delta}<\sigma\sqrt{2\epsilon\eta}
\end{equation}
which can be satisfied for sufficiently large detuning.
However, Eq. \ref{eq:condition2} for resolving single spin noise becomes
\begin{equation}
\frac{\sigma_r}{A}>\frac{1}{2\eta}
\end{equation}
which is likely an unphysical regime since
since $\sigma_r\sim \lambda^2$ where $\lambda$ is the laser wavelength and
$\eta$ is a small parameter.
This states that the photon shot noise for coherent light overwhelms the
single spin noise.  Using squeezed states of light with smaller quadrature
fluctuations may allow detection of single spin noise but a more detailed
analysis would be necessary.

\section{\label{sec:conclusion}Conclusion}
Ultracold atoms offer the prospect of experimentally realizing quantum many-body
systems in a strongly interacting regime with advancements such as
optical lattices, Feshbach resonances, and confinement to low dimensionality.
Noise analysis, or the study of fluctuations of an operator potentially
to high orders instead of just the average, offers an avenue to
access correlation functions and to characterize many-body states.
We discussed a proposal for using noise analysis in quantum spin systems realized
with ultracold atoms.  By studying the entire distribution function
for the magnetization of a mesoscopic spin system, the behavior of 
high order correlation functions can be extracted.  We analyzed the
Ising spin chain in a transverse field in detail, focusing on the transverse
magnetization.  The resulting distribution function possesses a distinct
splitting between even and odd values of the magnetization.  This
splitting  behaves differently in ordered, critical, and disorderd phases
and has a simple interpretation in terms of confinement versus proliferation
of domain walls.  Finally, we discussed pertinent issues for implementing
quantum noise analysis through experiments in the near future.

\acknowledgements{
We thank M. Lukin, E. Altman, A. Polkovnikov, L. Levitov, V. Gritsev 
for useful discussions.
This work was supported by NSF grant DMR-0132874, Harvard-MIT CUA, AFOSR, 
and NDSEG program.
}

\appendix
\renewcommand{\theequation}{\Alph{section}.\arabic{equation}}

\section{\label{app:toeplitz}Toeplitz Determinants}
In this appendix, we outline the derivation of the asymptotic form of the generating
function in Eq. \ref{eq:chin_asymptotic} using the theory of Toeplitz 
determinants.  Eq. \ref{eq:chin_f} gives an exact expression for 
the $\chi_n(m)$ as a determinant of a matrix with constant diagonals.
The matrix elements are given by the fourier coefficients $f_{i-j}(z)$ of 
the function $f(z,x)$ in Eq. \ref{eq:f}.
The theory of Toeplitz determinants (see Ref. \cite{bottcher-90}) 
gives the large-$n$ asymptotics in the form
of \ref{eq:chin_asymptotic}
with applications to a number of physical problems (see Ref. \cite{cherng-06} and
references therein).

The simplest case to consider is when $f(z,x)$ is continuous on $|x|=1$
with integer winding number where a variant of Sz\"{e}go's lemma holds 
\cite{bottcher-90}.
For power-law singularities in $f(z,x)$ (branch cuts in $\log f(z,x)$),
the Fisher-Hartwig conjecture is needed and has been proven in the cases
we need \cite{ehrhardt-01}.
These results are rigorous but we can give a heuristic motivation 
for their structure.
Formally, one can write
\begin{equation}
\det[f_{i-j}(z)]=\exp(\Tr[\log f_{i-j}(z)])
\end{equation}
where $i,j=1\ldots n$
and the leading contribution comes from the replacement
\begin{equation}
\Tr[\log f_{i-j}(z)]\rightarrow n
\int_{-\pi}^{\pi}\frac{dk}{2\pi} \log f(z,e^{ik}).
\end{equation}
This replacement corresponds to  noticing $f_{i-j}(z)$ is
translationally invariant in the sense that it depends only on $i-j$ 
and evaluating the trace in momentum space.
The subleading terms correspond to finite size corrections in $n$.

The functions $c_{i}(z)$ depend on the analytic structure of $\log f(z,x)$
and differ for $g<g_c$, $g=g_c$, and $g>g_c$.
It will be useful to define the following functionals acting on functions of $x$
\begin{eqnarray}
h_0[f(x)]&=&\frac{1}{2\pi i}\oint \frac{dx}{x} \log f(x)\\
h_y[f(x)]&=&-\frac{1}{2\pi i}\text{P}\oint dx \frac{y+x}{2x(y-x)}\log f(x)\\
E[f(x)]&=&\frac{1}{(2\pi i)^2}\text{P}\oint dx \oint dx' 
\frac{x+x'}{2x'(x-x')}\frac{f'(x)}{f(x)}\log f(x')
\end{eqnarray}
where the contour integrals are along the unit circle and notice $h_y[f(x)]$ depends
on an additional parameter $y$.  If there are singularities on the unit circle
then the integrals are in the principal value sense.

First we consider $g>g_c$.  For all $z$, $f(z,x)$ is continuous 
on $|x|=1$ with zero winding number.  We obtain
\begin{eqnarray}
\label{eq:asymptotic_above}
c_0(z)&=&h_0[f(z,x)]\\
c_1(z)&=&0\\
c_2(z)&=&E[f(z,x)].
\end{eqnarray}

Next we consider $g<g_c$.  For $\MyRe[z]\ge0$ ($\MyRe[z]<0$), 
$\log f(z,x)$ is non-singular with winding number zero (one).
We obtain
\begin{eqnarray}
\label{eq:asymptotic_below}
c_0(z)&=&\begin{cases}
h_0[f(z,x)]&\MyRe[z]\ge0\\
h_0[x^{-1} f(z,x)]+\log(-\xi)&\MyRe[z]<0
\end{cases}\\
c_1(z)&=&0\\
c_2(z)&=&\begin{cases}
E[f(z,x)]&\MyRe[z]\ge0\\
E[x^{-1}f(z,x)]+2 h_\xi[x^{-1}f(z,x)]-\log \frac{df}{dx}(z,\xi)&\MyRe[z]<0
\end{cases}
\end{eqnarray}
where $|\xi|<1$ corresponds to the singularity of $\log f(z,x)$ 
closest to the unit circle given by
\begin{equation}
\label{eq:xi}
\xi=\frac{2g+i\sqrt{2-z^2-z^{-2}-4g^2}}{z+z^{-1}-2}.
\end{equation}
and the square root is positive for positive real arugments.

The final case we consider is $g=g_c$.  Here, $f(z,x)$ is discontinuous
at $x=-1$ for all $z$.  This singularity can be written in power-law form
\begin{equation}
g(z,x)=\left(\frac{1+x}{1+x^{-1}}\right)^{\alpha(z)}
\end{equation}
where the exponent is given by
\begin{equation}
\alpha(z)=\frac{1}{4}-\frac{1}{\pi}\arctan(z).
\end{equation}
Notice $\log f(z,x)/g(z,x)$ is continuous.  We obtain
\begin{eqnarray}
\label{eq:asymptotic_crit}
c_0(z)&=&h_0[f(z,x)/g(z,x)]\\
c_1(z)&=&-\alpha(z)^2\\
c_2(z)&=&E[f(z,x)/g(z,x)]+2\alpha(z) h_{-1}[f(z,x)/g(z,x)]+
\log[G(1+\alpha(z))G(1-\alpha(z))]
\end{eqnarray}
where $G(x)$ is the Barnes-$G$ function satisfying $G(x+1)=\Gamma(x)G(x+1)$
with $\Gamma(x)$ the gamma function.

\section{\label{app:g}Small and large $g$ limits}
We now turn to the small and large $g$ expansions of the generating function
$\chi_n(\lambda)$ in Eq. \ref{eq:chin_g}.
For $g\ll g_c$, we use the determinant representation in Eq. \ref{eq:chin_f}
and first expand the matrix elements $f_{n}(z)$ in powers of $g$.  We find
\begin{equation}
f_n(z)=
\begin{cases}
\frac{2+g}{4}+\frac{2-g}{4}z+O(g^2)& n=0\\
+g\frac{1-z}{2}+O(g^2)& n=1\\
-g\frac{1-z}{2}+O(g^2)&n=2\\
O(g^2)&n\ne 0,1,2
\end{cases}.
\end{equation} 
The determinant $\det[f_{i-j}(z)]$ is lower triangular and given by 
the product of the diagonal matrix elements 
\begin{equation}
\chi_n(\lambda)=\left(\frac{2+g}{4}+\frac{2-g}{4}e^{i\lambda}\right)^n.
\end{equation}
as in Eq. \ref{eq:chin_g}.  This corresponds to the generating function for
a binomial random variable and the identification of $\rho_n\sim B(n,(2-g)/4)$ 
in Eq. \ref{eq:rhon_g}.
A binomial variable $B(n,p)$ with $n$ trials and probability of success $p$
has a distribution
\begin{equation}
\label{eq:binomial}
\Prob[B(n,p)=m]=\frac{\Gamma(n+1)}{\Gamma(m+1)\Gamma(n-m+1)}(1-p)^{n-m}p^m
\end{equation}
where $\Gamma(n)$ is the gamma function.
The distribution $P_n(m)$ is given by Eq. \ref{eq:binomial} with $p=(2-g)/4$
which we use to plot Fig. \ref{fig:compare}.

For $g\gg g_c$, we analyze the determinant of Eq. \ref{eq:chin_g}
using the large $n$ asymptotic of Eq. \ref{eq:chin_asymptotic} as discussed
in Appendix \ref{app:toeplitz}.
Instead of the integral representation employed in Eq. \ref{eq:asymptotic_above}
it will be easier to use the equivalent representation 
\begin{equation}
\chi_n(\lambda)\sim \exp\left[h_0(z) n +\sum_{k=1}^{\infty} kh_k(z)h_{-k}(z)\right]
\end{equation}
in terms of the fourier coefficients of $\log f(z,x)$ given by
\begin{equation}
\label{eq:chin_h}
\log f(z,x) = \sum_{n=-\infty}^{\infty} h_n(z) x^n.
\end{equation}
We expand the coefficients $h_n(z)$ in powers of $g$ and find
\begin{equation}
h_n(z)=
\begin{cases}
\frac{1}{16g^2}(z^2-1)+O(g^{-3})&n=0\\
\mp \frac{1}{4g}(z-1)+O(g^{-2})&n=\pm 1\\
O(g^{-3})&n\ne 0,\pm 1
\end{cases}
\end{equation}
Using Eq. \ref{eq:chin_h}, this gives the generating function to $O(g^{-3})$ as
\begin{equation}
\chi_n(\lambda)=\exp\left[\frac{n-1}{16 g^2}(e^{2i\lambda}-1)+\frac{1}{8 g^2}(e^{i\lambda}-1)\right]
\end{equation}
which is that of the sum of two Poissonian random variables as in Eq. \ref{eq:chin_g}.
However, notice one term enters with $e^{2i\lambda}$ and the other enters with 
$e^{i\lambda}$ meaning we identify $\rho_n\sim 2 P((n-1)/16g^2)+P(1/8g^2)$
in Eq. \ref{eq:rhon_g}.
A Poissonian variable $P(\lambda)$ with parameter $\lambda$ has a distribution
\begin{equation}
\Prob[P(\lambda)=m]=\frac{e^{-\lambda}\lambda^m}{\Gamma(m+1)}
\end{equation}
and a sum of two Poissonian variables of the form
\begin{equation}
X=2P(\lambda_2)+P(\lambda_1)
\end{equation}
has a different distribution for even and odd values
\begin{eqnarray}
\Prob[X=2m]&=&\sum_{n=0}^{m}\Prob[P(\lambda_2)=n]\Prob[P(\lambda_1)=2(m-n)]\\
\Prob[X=2m+1]&=&\sum_{n=0}^{m}\Prob[P(\lambda_2)=n]\Prob[P(\lambda_1)=2(m-n)+1].
\end{eqnarray}
These sums can be evaluated in terms of the hypergeometric function ${_1F_1}(a;b;z)$
\begin{equation}
\label{eq:poisson}
\Prob[X=m]=\exp\left[-\frac{(2\lambda_2+\lambda_1)^2}{4\lambda_2}\right]\lambda_2^{m/2}
\begin{cases}
\frac{1}{\Gamma\left(\frac{m+2}{2}\right)}{_1F_1}\left(\frac{m+1}{2};\frac{1}{2};\frac{\lambda_1^2}{4\lambda_2}\right)
& m\ \text{even}\\
\frac{1}{\Gamma\left(\frac{m+1}{2}\right)}\frac{\lambda_1}{\sqrt{\lambda_2}}{_1F_1}\left(\frac{m+2}{2};\frac{3}{2};\frac{\lambda_1^2}{4\lambda_2}\right)
&m\ \text{odd}
\end{cases}
\end{equation}
The distribution $P_n(m)$ is 
given by Eq. \ref{eq:poisson}
with $\lambda_1=1/8g^2$, $\lambda_2=(n-1)/16g^2$
which we use to plot Fig. \ref{fig:compare}.

\section{\label{app:saddlepoint}Saddle point analysis}
In this appendix, we obtain the distribution $P_n(m)$ by evaluating the
integral of Eq. \ref{eq:pn} in the saddle-point approximation.
We rewrite the integral as a contour integral over the unit circle
$|z|=1$
with the asymptotic form of $\chi_n(\lambda)$ in Eq. \ref{eq:chin_asymptotic}
to obtain
\begin{equation}
\label{eq:pn_contour}
P_n(m)\sim\frac{1}{2\pi i}\oint \frac{dz}{z^{m+1}}\exp\left[c_0(z) n + 
c_1(z) \log n +c_2(z)\right].
\end{equation}
The leading contribution to integral for large $n$ is controlled by the
term $z^{-m}\exp[c_0(z)n]$ which has saddle points at
\begin{equation}
\label{eq:sp}
m=nzc_0'(z)
\end{equation}
where primes denote differentiation with respect to $z$.  For the function
$c_0(z)$ obtained in Appendix \ref{app:toeplitz}, we find there
are two saddle points.  One is located on the positive real $z$ axis and the other
is on the negative real $z$ axis which we denote as $z_\pm$ respectively.
In addition, the $|z|=1$ contours can be deformed to steepest descent contours and
we obtain in the saddle-point approximation
\begin{equation}
\label{eq:pn_saddlepoint}
P_n(m)\sim\sum_\pm \pm \frac{z_\pm^{-(m+1)}
\exp\left[c_0(z_\pm) n+ c_1(z_\pm)\log n+c_2(z_\pm)\right]}
{\sqrt{2\pi[(m+1)/z_\pm^2+n c_0''(z_\pm)]}}.
\end{equation}

We have already argued that the in the large-$n$ limit, $P_n(m)$ should
be gaussian in the discussion of Eq. \ref{eq:gaussian}.  In general, the
gaussian form of $P_n(m)$ is good near the peak of the distribution.
Thus, we will perform an expansion of Eq. \ref{eq:pn_saddlepoint}
about the peak located at $n\mu_\pm$ and the corresponding 
saddle-points at $z^0_\pm$.
The two quantities are related by the zeroth order saddle-point condition
\begin{equation}
\label{eq:mu}
\mu=z^0 c_0'(z^0) 
\end{equation}
and we have dropped the $\pm$ subscript for clarity.
We then substitute
\begin{eqnarray}
m&=&n\mu+n\delta m\\
z&=&z^0+\delta z
\end{eqnarray}
into the full saddle-point equation of Eq. \ref{eq:sp}
and solve order by order to obtain
\begin{equation}
\delta z=a_1 \delta m + a_2 \delta m^2
\end{equation}
where the coefficients $a_i$ are given by
\begin{eqnarray}
a_1&=&\frac{1}{c'_0(z^0)+z^0c''_0(z^0)}\\
a_2&=&-\frac{2c''_0(z^0)+z^0c'''_0(z^0)}
{2[c'_0(z^0)+z^0c''_0(z^0)]^3}.
\end{eqnarray}

Evaluating Eq. \ref{eq:pn_saddlepoint} up to second order
in $\delta m$ for the leading term $z^{-m}\exp[c_0(z)n]$ and 
zeroth order for the other terms yields
\begin{equation}
\label{eq:pn_almost_gaussian}
P_n(m)\sim \sum_\pm \frac{\pm 1}{z^0_\pm} 
\frac{e^{-(m-n\mu_\pm)^2/2n\sigma^2_\pm-m\log z^0_\pm}}{\sqrt{2\pi n\sigma^2_\pm}}
\exp\left[c_0(z^0_\pm)n + c_1(z^0_\pm)\log n + c_2(z^0_\pm)\right]
\end{equation}
where $\sigma^2_\pm$ is given by
\begin{equation}
\label{eq:sigma}
\sigma^2=z^0[c'_0(z^0)+z^0c''_0(z^0)]
\end{equation}
and the $\pm$ subscripts have been suppressed.
Since $n\mu$ should correspond to the peak of the distribution, we see
that $z^0$ is determined by the condition that the term linear in $m$ of
the exponent in Eq. \ref{eq:pn_almost_gaussian} should have a real part of zero
to ensure that it does not shift the peak.
This implies that $z^0_\pm=\pm 1$.

At this point, we have to check whether the $z^0_\pm$ saddle-points
contribute to the leading term $z^{-m}\exp[c_0(z)n]$ at the same order.
The functions $c_0(z)$ are given in Appendix \ref{app:toeplitz} and 
from Eq. \ref{eq:asymptotic_above} and Eq. \ref{eq:asymptotic_crit} we find
that $c_0(\pm 1)=0$ which means that both saddle-points should be included for $g\ge g_c$.
However, from Eq. \ref{eq:asymptotic_below}, we see that $c_0(+1)>c_0(-1)$ 
due to the additional contribution of $\log(-\xi)$ of Eq. \ref{eq:xi} with
$|\xi|<1$.  Thus the contribution from $z^0_-$ should be dropped for $g<g_c$.  

Evaluating the remaining
expressions in Eq. \ref{eq:pn_almost_gaussian} using the functions $c_i(z)$ obtained
in Appendix \ref{app:toeplitz} gives the gaussian form of Eq. 
\ref{eq:pn_gaussian}.  In addition we find that $\mu_\pm=\mu$ given by
Eq. \ref{eq:mu} corresponds to the
mean and $\sigma_\pm^2=\sigma^2$ given by Eq. \ref{eq:sigma}
corresponds to the variance and are given
by the expressions in Eq. \ref{eq:pn_parameters}.
An enlightening way of writing Eq. \ref{eq:pn_almost_gaussian} is given by
\begin{equation}
\label{eq:pn_alt}
P_n(m)\sim \frac{e^{-(m-n\mu)^2/2n\sigma^2}}{\sqrt{2\pi n\sigma^2}}\left[\chi_n(0)+(-1)^m\chi_n(\pi)\right].
\end{equation}
where $\chi_n(\lambda)$ with $\lambda=0,\pi$ correspond to contributions from the zeroth order
saddle points $z=e^{i\lambda}=+1,-1$,
respectively.
The contribution from the $z=+1$ saddle point is dictated by normalization of the distribution function 
\begin{equation}
\chi_n(0)=\sum_m P_n(m)=1.
\end{equation}
However, Eq. \ref{eq:tautau_identity} shows the $z=-1$ contribution 
\begin{equation}
\chi_n(\pi)=\sum_m P_n(m) (-1)^m=4\langle T_x(0)T_x(n)\rangle.
\end{equation}
is determined by the $T_x(0)T_x(n)$ correlation function.
Since the asymptotic expression we obtained for $\chi_n(\lambda)$ is exact for any $\lambda$
and the even versus odd splitting is given precisely by $\chi_n(\pi)$, we see that the splitting
gives the exact asymptotic behavior of $\langle T_x(0)T_x(n) \rangle$.

\bibliography{references}
\bibliographystyle{apsrev}

\end{document}